# A Study on Mars Probe Failures

Malaya Kumar Biswal M[*]  and Ramesh Naidu Annavarapu[†]

*Department of Physics,
School of Physical Chemical and Applied Sciences,
Pondicherry University, Kalapet, Puducherry, India – 605 014*

The long term dreams to approach Mars requires numerous spacecraft attempts for exploration as well as to understand the perception of the red planet. Before launching a mission, the space probe undergoes critical ground testing and effective preparation. Though probes were carefully tested and validated, many experiences temporary or permanent setbacks prior to their final state of mission accomplishment, resulting in the failure of the mission. In order to figure out the problems concerning probe malfunction or failure, we conducted a study on failed Mars probes that are launched between 1960 to 2020. The probes were characterized to determine various modes of failure and their impact on the missions. The results of our study from past probes showed effective integration and testing, sterling fabrication and validation of space probes, adequate software design, feasible recovery options, and novel guidance to probe computers and communication systems.

## I. Nomenclature

| | | |
|---|---|---|
| *EDL* | = | Entry, Descent, and Landing |
| *EDM* | = | Entry, Descent, Module |
| *FCP* | = | Failure in Computer Programming |
| *FCS* | = | Failure of Communication System |
| *FDEIS* | = | Failure of Descent Engine Ignition System |
| *FIS* | = | Failure of Ignition System |
| *FOCS* | = | Failure of Onboard Computer System |
| *FOS* | = | Failure of Orientation System |
| *FSF* | = | Failure of Spacecraft Function |
| *FSP* | = | Failure of Software Program |
| *FTCS* | = | Failure of Thermal Control System |
| *JAXA* | = | Japanese Aerospace Exploration Agency |
| *LBMI* | = | Lost Before Mission Initiation |
| *LEO* | = | Low Earth Orbit |
| *LVM* | = | Launch Vehicle Malfunction |
| *MV* | = | Mars Venus |
| *USA* | = | United States of America |
| *USSR* | = | Union Soviet Socialist Republic |

## II. Introduction

Global Space Communities have launched 72 Mars probes (includes probes and sub-probes) in 47 attempts. Unfortunately, 33 probes encountered setbacks. Because the red planet is too far to reach together with a challenging environment. Despite challenges, we have reached out Mars with 33 successful probes and 5 en-route probes [7-9]. Similarly, 33 probes encountered failure before accomplishing their mission intent, and out of 33 failed probes, 9 probes were never deployed before its mission commences. Hence, to figure out the possible causes behind their failures, we made a study on 33 identified malfunctioned probes that are launched within the time frame of 1960 and 2020. The reports were gathered from [1-6] and are verified with official reports released by the mishap investigation board commissioned by respective space agencies. The reports were carefully analyzed to distinguish the various mode of probe failures and their malfunctioned components. We have also studied the stages of recurrent failures of Mars probes and their components. Further, our analysis compares different sorts of failure systems and probe parameters to determine their impact on the mission. Our study is completely current and is different from other reports especially made on Mars probes. Moreover, technical and descriptive report on Mars probe failures can be found at Mars Failure Report [1].

---

[*] Graduate Researcher, Department of Physics, Pondicherry University, India; malaykumar1997@gmail.com, mkumar97.res@pondiuni.edu.in, Student Member of Indian Science Congress Association, Student Member AIAA.
[†] Associate Professor, Department of Physics, Pondicherry University, India; rameshnaidu.phy@pondiuni.edu.in. Non-Member AIAA.

### III. Classifications and Research Methodology

**A. Classifications**

For our study, we have classified various modes failures into the following categories: Launch Vehicle Malfunction (LVM) which includes issues related to launchers, booster stage fire accidents, and explosions; Failure of the Ignition System (FIS) which is the combination of failure of ignition system and stage separation payload fairing; Failure of Communication System (FCS) deals with the issue concerns with antenna and coronal discharge; Failure of the Orientation System (FOS) that includes the orientation of the probes and solar panels; Failure of Thermal Control System (FTCS) includes issues concerning the imbalance in internal temperature and thermal maintenance of the probe; Failure of the Onboard Computer System (FOCS) is associated with the failure of on-computer channels and concurrent reboots; Failure of Software Program (FSP) deals with software development and flaws; Failure in Computer Programming (FCP) includes the wrong programming and wrong command sent to probes without proofreading and validation by humans; Failure of Descent Engine Ignition (FDEIS) is the obstacle in firing the breaking engine for Mars atmospheric entry and EDL descent; Failure of Spacecraft Function (FSF) deals with the unexpected malfunction of the whole probe and finally Lost Before Mission Initiation (LBMI) is associated with the probes that were never deployed from the main bus or lost with the launch vehicles and the probes that stranded in interplanetary space.

**B. Research Methodology**

- Our study gathers failure reports of Mars probes from [1-6] and appropriate online resources of space agencies.
- We found that the failure was caused as a result undesirable events such as of a launch failure, probes that are stranding in low-earth orbit or destroyed during re-entry, sailing on the trajectory phase to Mars, probes that missed the planet due to unsuccessful Mars orbital insertion, and probes that crashed into the surface as a consequence to complete malfunction.
- We identified the first source of failure that triggers subsequent issues within the probes and were clearly depicted as stages of component failures in table 1.
- Then, based on the acquired report, we defragmented the spacecraft or probe into discrete segments shown in Chart 1 and Chart 2 and categorized under the classification discussed in "Classifications".
- Distinct components of each Mars probes were counted and enlisted as shown in [10]. In addition to this, Mars probe's properties like power, dry mass, launch mass, time duration after which the probe encounters its first source of failure encounter, mission duration and mission degradation were comprehensively collected. Finally, their proportion is estimated.
- Our study provides a brief outline to understand the possible causes responsible for the failure of Mars probes that were launched between 1960 and 1973. Since many space agencies never revealed the failure report and the technical report on probe specification as part of their secret agenda. But acceptable causes for the failure of Mars probes were briefly reported in [1].
- Here in our study, we have excluded the en-route probes that are launched during the 2020 Mars window because these probes were supposed to orbit or land on Mars in late February 2021. And it is hard to interpret their mission status whether it may encounter success or failure.

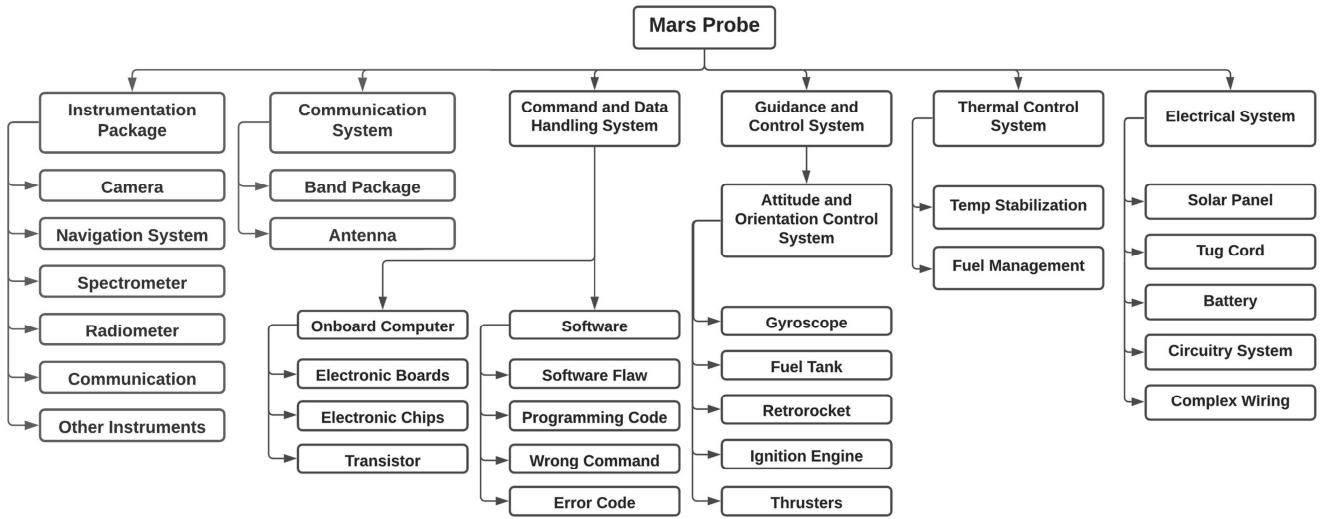

**Chart 1 Anatomy of a Defragmented Mars Probe**

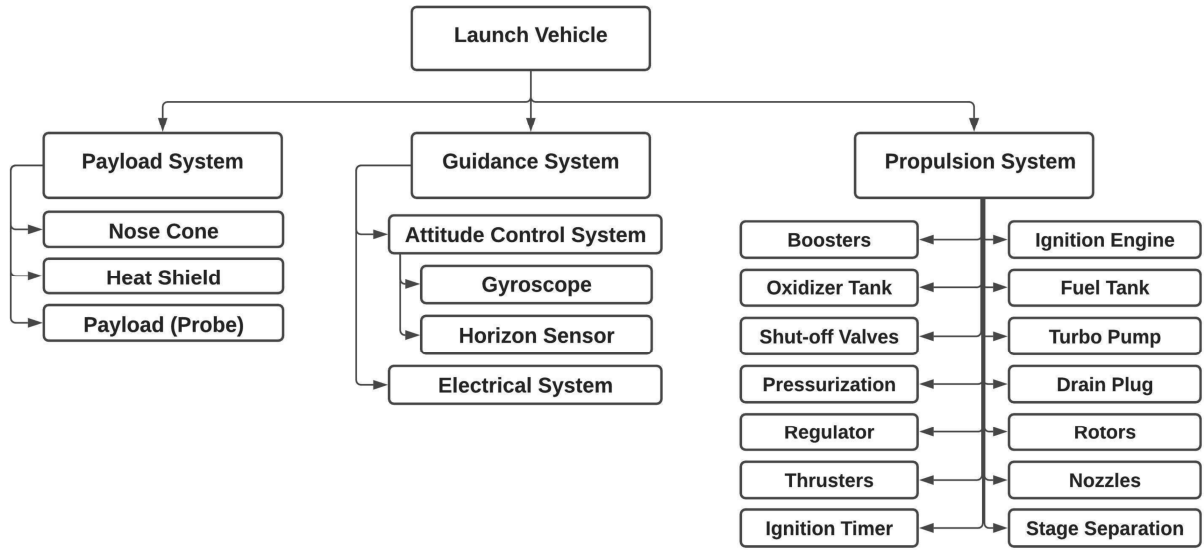

**Chart 2 Anatomy of Defragmented Launch Vehicle**

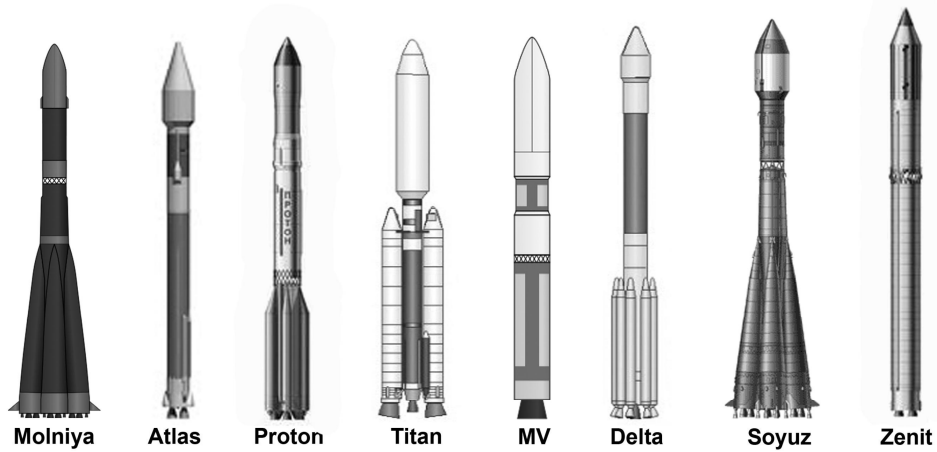

**Launch Vehicle Considered for Defragmentation**

3
American Institute of Aeronautics and Astronautics

## IV. Overall Failure Analysis

### A. Success vs. Failure Proportion

This section presents an outline of overall missions and their failure proportion. Fig.1 shows the proportions of failure in terms of mission attempts and Mars Probe counts. The failure proportion of the number of probes is equal to the number of probes achieved success, but the failure rate is fairly higher than the success rate in terms of mission attempts. Thus, this drives the attention to consider and conduct failure analysis on Mars probes. In addition to this, we have excluded the en-route probes like Mars Perseverance Rover, Emirates Mars Mission, and Tianwen-1. Because it is uncertain to predict their mission status as the probes are likely to approach Mars in February 2021.

### B. Nations Failure Proportion

We have also estimated failure proportion by nation shown in Fig 3. An overall estimate shows that the Soviet Union (USSR) is found to have more probe loss (71%) than the United States (17%), and the European Union (6%). China and Japan have the same proportion of probe loss (3%). Besides the United Arab Emirates have attempted Emirates Mars Mission, China's Tianwen-1, and United States (USA) Mars 2020, these probes are in trajectory cruise to Mars.

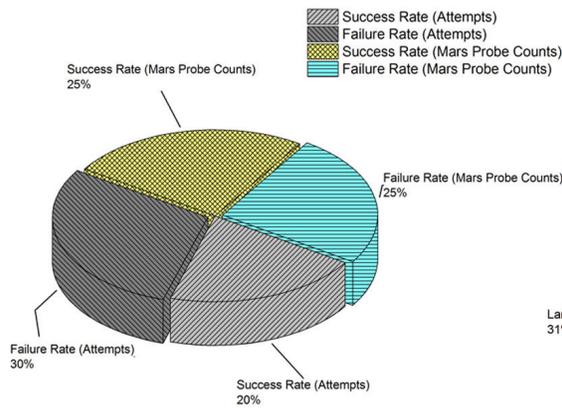

**Fig 1 Success vs Failure Proportion**

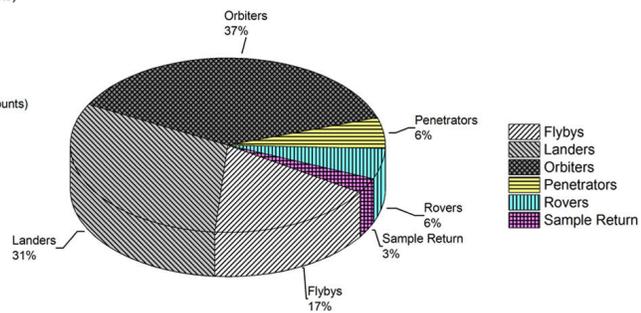

**Fig 2 Failure Proportion of Mars Probe Types**

### C. Failure Proportion by Probe Types

An analysis over probe of distinct type showed that Orbiters have the highest proportion of failure rate (37%) as compared to the Landers which is (31%), Flybys (17%), and Sample Return Vehicles (3%). Penetrators (Impactors) and Rovers have equal failure proportion that is (6%). The most failure rate of Orbiters reflects that most tragedy occurs at Mars orbit than the surface. Overall failure estimate is shown in Fig 2.

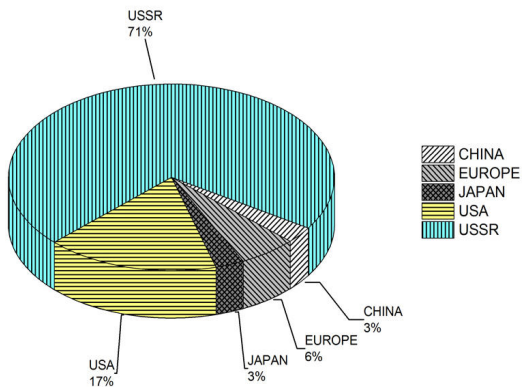

**Fig 3 Nations Failure Proportion**

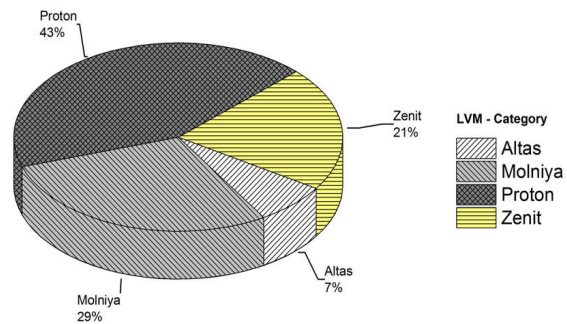

**Fig 4 Failure Proportion by Launchers**

### D. Failure Proportion by Launchers

Proportional analysis over launchers showed that (43%) probes lost with Russian's Proton Rocket that stands to be the first major than Molniya (29%). Then 21% of probes lost with Russian's Zenit launcher and 7% with United States Atlas launch vehicle. Most of the launch vehicle issues were enhanced nowadays, but the recent launcher attempt in 2011 to launch the Fobos-Grunt mission persisted this kind of issue. Hence, it is recommended to have proper technical attention for testing and effective integration to avoid launcher failures in future. Overall failure proportion is shown in Fig 4.

### E. Failure Proportion by Classification

We have classified overall failure types into eleven categories that enclosed 33 Mars probes launched in 26 attempts. Fig 5 shows the classification of various failure modes. We observe that 26% of Mars probes seem to have lost before mission commencement (i.e. the probe that either lost with orbiter's main bus or destroyed). Similarly, 14% of probes were found to have encountered with issues concerning ignition engine or system (i.e. either in stage separation or payload firing or engine plighted for mid-course corrections). Following, 11% affects the launch vehicles and fire accidents of strap-on boosters. In addition to this, from a technical perspective, we observe that most of the failure affects communication systems and spacecraft software whose failure proportion is found to be 9%. As well 6% of all failures contribute to the failure of thermal and orientation control system. Finally, the failure of the descent engine (allocated for supporting the Mars EDL event) and unanticipated complete malfunction of probe contribute 3% which is least of all failures. Hence, we have to concentrate mainly on the technical section of the Mars probe (i.e. development of robust software, adequate programming and proofreading procedures, effective design of orientation control system, and perfect thermal insulation of Mars probes).

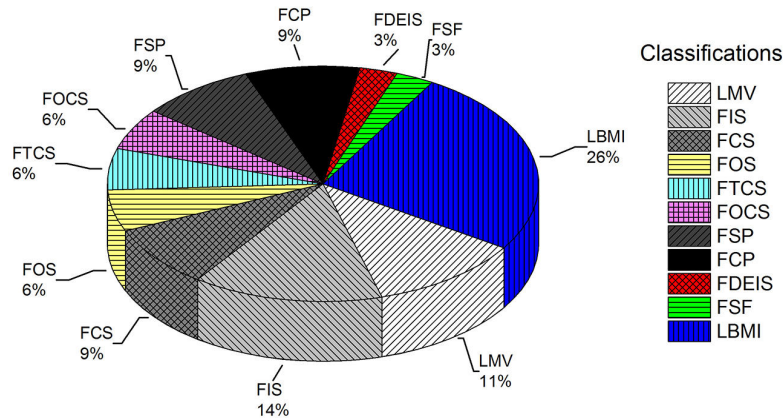

**Fig 5   Proportion of Classified Mars Probe Failures.**

### F. Failure Proportion of Malfunctioned Probe Components

In this section, we have performed a proportional analysis in terms of Mars probe's components from spacecraft fragments shown in Fig 6 and Fig 7, where Fig 6 shows the total number of components in terms of counts from overall analysis and their corresponding proportion in Fig 7. We observe that most affected components are the communication system and the ignition engine that contributes (14%) of all components and it remains a probable cause for the failure of most Mars probes (shown in Fig 5). The second most malfunctioned component is the Solar Panel (6%) due to their concerns in deploying and orientation. Consequently, the same (6%) affects onboard computer whose failure was attributed to the defective transistor (5%). Similar to the transistor, the probe's battery contributes (5%) of all components. Further, (3%) of components influences attitude control system, electrical system, orientation control system, software flaw and programming, and turbopump. Finally, the failure proportion (1%) of all components is attributed to descent engine, drain plug, electronic chips, fuel tank rupture, heat shield, horizon sensor, fuel leakage, oxidizer pump, oxidizer shut-off valve, parachute, pressurization regulator, pressurization system, pyro valve, retrorocket, rotor, the whole probe, stabilization system, and tug cord.

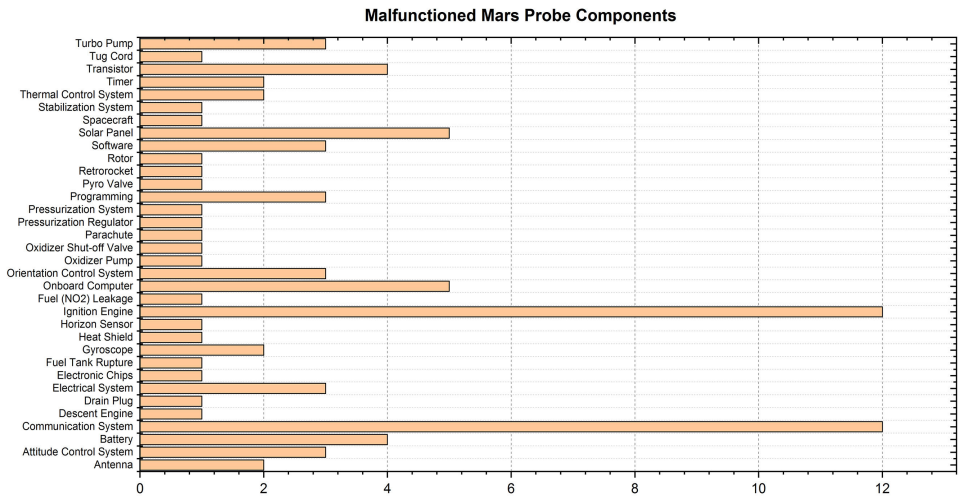

**Fig 6  Malfunctioned Probe Components**

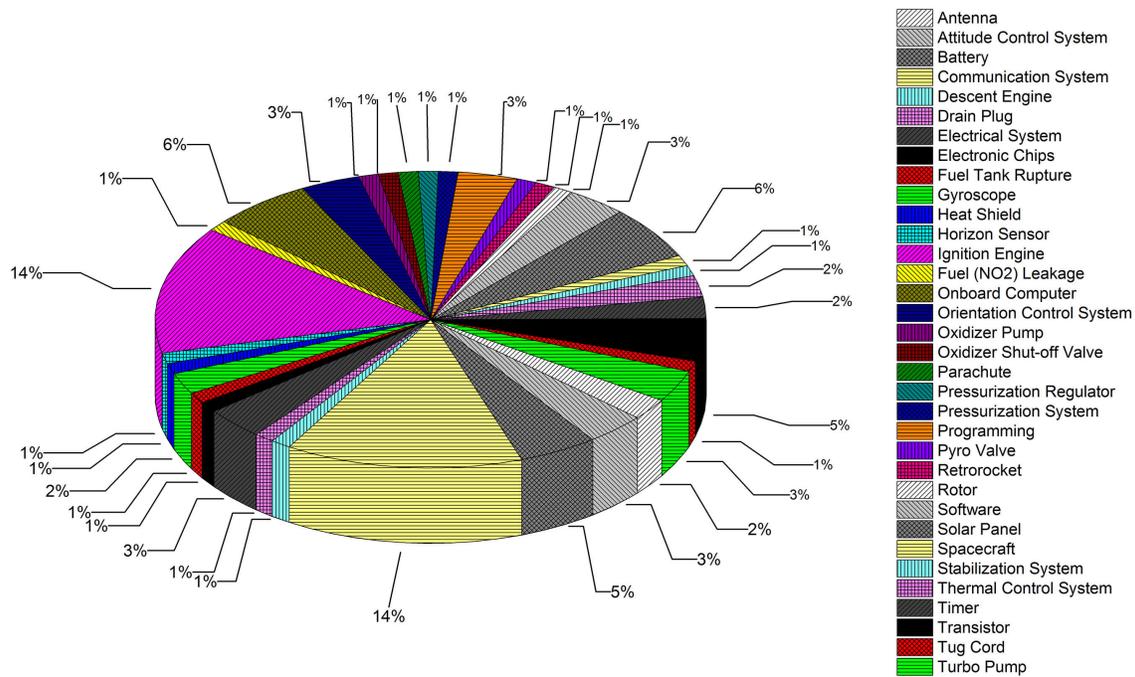

**Fig 7  Proportion of Malfunctioned Mars Probe Components**

## V.  Classified Failure Analysis

### A.  LVM failure analysis

    We see that the second most failure occurs due to the malfunction of launch vehicles (11%) shown in Fig 5. The probes fall under this category are Mars 1M No.1 whose launch vehicle malfunction was triggered by a faulty gyroscope and its resonant vibration; Mars 2MV-4. No.1 whose launcher persisted lubricant leakage from turbo-pump resulting in the explosion of boosters; Mars 2M.No.522 has issues with leakage of NO$_2$ due to lack of drain plug; and Mariner 8 whose launcher circuitry depleted due to defective transistor. Finally, our analysis shows that the failure was condemned to inadequate launcher integration, thrusters testing and validation. Mostly the failure was directed to the malfunction of components such as turbo-pump, gyroscope, launcher circuitry, and drain plug.

### B. FIS failure analysis

We observe that a large number of failures occur in the FIS category. The ignition engine encounters firing issues and their failure was attributed to improper fuel pressurization, fuel leakage, and impairment of propeller's rotors. Notable probes include Mars 1M. No.2, 2MV-3. No. 1, Mariner 3, 2M.No.521, and Mars 96 orbiter. This signifies inadequate vehicle design from manufacturers and it contributes 14% of all failures. Hence, it is substantial to focus on this issue for a better insight to upcoming missions.

### C. FCS failure analysis

Impairment of communication system contributes 9% of all failures. In our analysis, we have found that the probes upon landing on the Mars surface counteract communication issue against rough terrain leading to coronal discharge. Noticeable probes were Mars 3 and Mars 6 lander. In addition to this, one of the probe (Mars Observer) whose communication system was affected by the lack of power that ultimately affected by the disoriented solar panel. Therefore, this issue can be minimized with the effective design of terrain proof antennas and communication systems.

### D. FOS failure analysis

Failure of the orientation system contributes 6% of all failures. Past incidences include: improper fuel leakage from Mars 2MV-4 No.2 probe accelerated the spinning movement of the probe resulting in the failure of probe orientation. Similarly, the damage of onboard computer channel of Phobos 2 orbiter adversely affected the orientation control leading to the complete loss of control over orientation. So, Improper orientation of probe may advert-off antenna from directed communication from the Earth and may disorient solar arrays from the Sun affecting the probe's power production. Hence, these issues can be addressed with proper probe configuration with a robust computer.

### E. FTCS failure analysis

The thermal stability and internal temperature of the probe are significant for fuel management and to determine the reliability of the spacecraft component. Probes under this category are found to have encountered this issue during interplanetary injection and interplanetary transit to Mars. The failure encounter was due to unsuited and complex space environmental condition [11]. FICS contribute 6% of all failures, noticeable probes are Zond 2 and Nozomi.

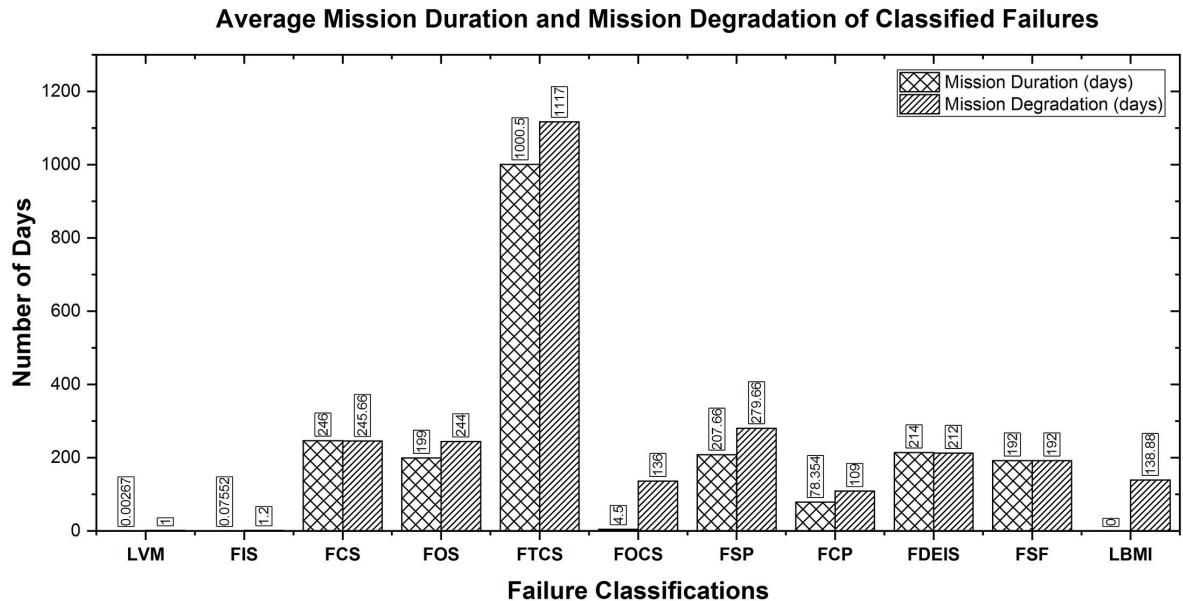

**Fig 8   Distribution of Mission Duration and Degradation of Classified Failures**

**F. FOCS failure analysis**

The Onboard computer of the probe plays a significant role in controlling the probe's subsystem and components. In our study, FOCS constitutes 6% of all failures and it was due to either malfunction of onboard computer channel or depletion of circuitry system as a result of a defective transistor. Noticeable probes were Mars 4 orbiter and Fobos-Grunt.

**G. FSP and FCP Failure Analysis**

FSP and FCP contribute 9% of all issues. The problem with FSP and FCP ensues as a result of flawed computer software and wrong programming. Our study has found six probes that have a backlog with this kind of issue. Observable probes were Mars Climate Orbiter, Mars Polar Lander, Schiaparelli EDM, Kosmos 419 Orbiter, Phobos 1 Orbiter, and Beagle 2 Lander.

**H. FDEIS and FSF failure analysis**

FDEIS and FSF constitutes (3%) of all failures. FDEIS occurs due to malfunction of descent engine allocated for firing during Mars atmospheric entry. Similarly, FSF occurs due to unexpected malfunction of the complete probe. Notable probes are Mars 7 lander (its descent engine failed to fire during hypersonic entry) and Mars 2 lander that malfunctioned before reaching Mars. Further, the distribution of probe mass, launch mass, mission duration, and mission degradation is shown in Fig 8 and Fig 9.

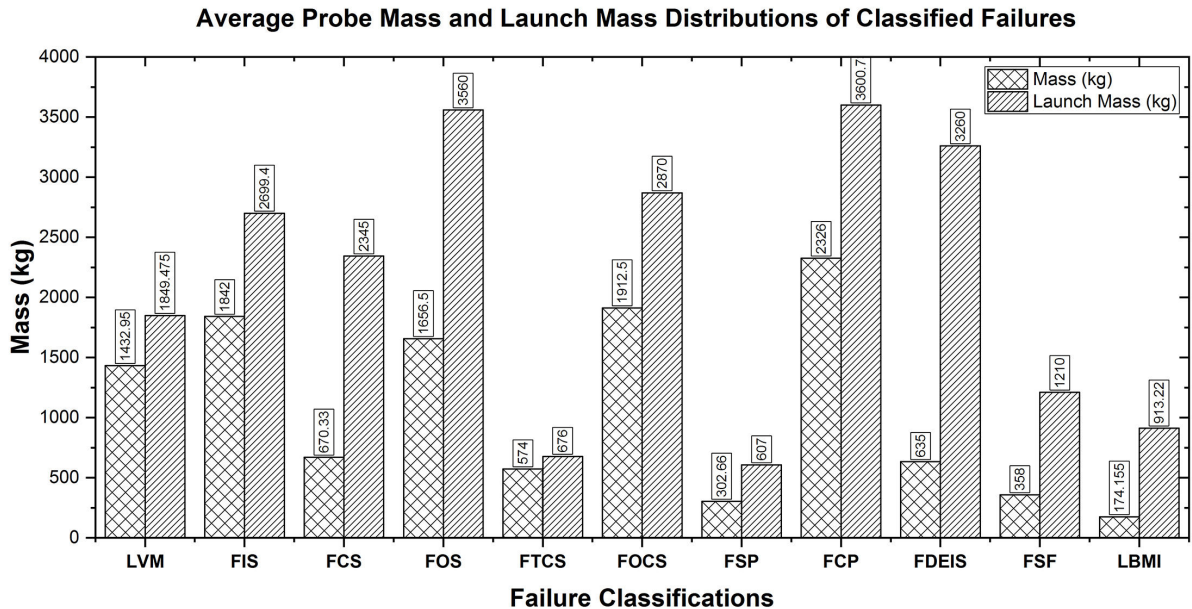

Fig 9 Distribution of Probe and Launch Mass of Classified Failures

## VI. Power Distribution and Time Duration of First Failure Encounter

**A. Power Distribution of Mars Probes**

Power distribution is estimated for overall failure categories. We have found that (11%) of Mars probes have power generation or power limit of (0-100 watts), 14% of probes have (100-500 watts), 3% have (500-1000 watts), and the same (3%) have (1000-2000 watts). Overall power distribution is shown in Fig 10. Further, we observed that the power distribution of (69%) of Mars probes is unspecified because many space firms never revealed their mission strategy as part of their secret agenda. Our study found most of the unknown parameters of space probes were found among Russian probes [12].

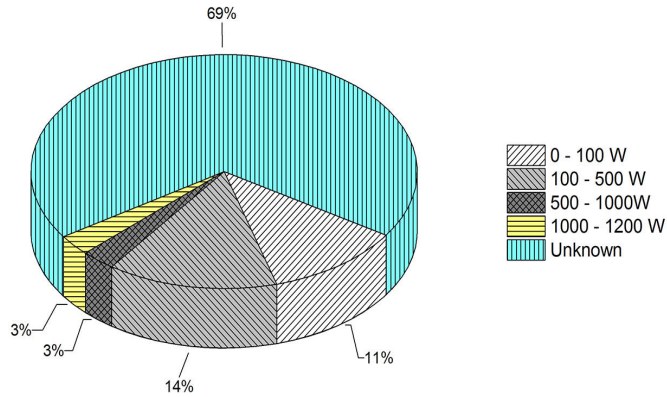

**Fig 10 Proportion of Power Distribution of Mars Probes**

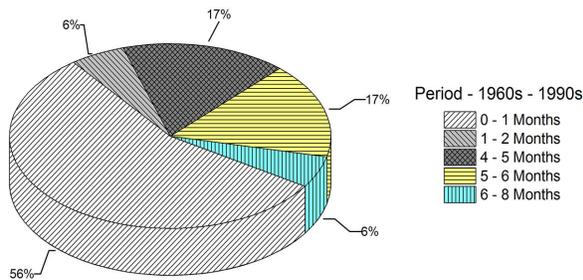
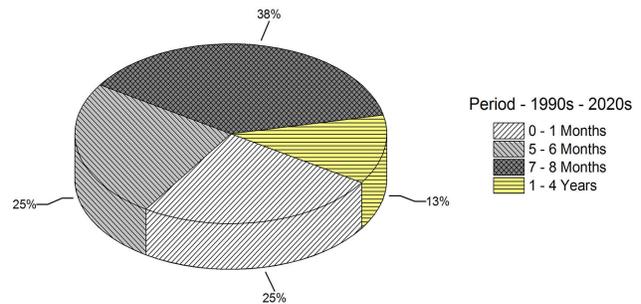

**Fig 11 Time Duration (1960s – 1990s)**   **Fig 12 Time Duration (1990s – 2020s)**

### B. Time of Failure Encounter

In this section, we have estimated the time duration after which the probe encounters its first source of failure after launch. So we have divided the time frame into two segments. One frame is between the period from the 1960s to 1990s and another time frame is from 1990s to 2020s.

**Time Frame - 1960s to 1990s**

The probes that are launched between 1960 and 1990 is found to have the first source of failure encounter after (0-1 months) at the proportion of (56%), then (6%) probes have failure issues after (1-2 months), (17%) probes have after (4-5 months), and the same (17%) probes have failure effect after (5-6 months), finally (6%) of all probes have after an operational period of (6-8 months). Comparably the probes that have the first source of failure within (0-1 months) are due to repeated launch failures.

**Time Frame - 1990s to 2020s**

Here the time duration increases and most of the probe encounters its first source of failure encounter after (7-8 months) with a proportion of (33%) and (13%) probes encounter after (7 months to 4 years). It clearly pictures that the failure rate of the Mars probe decreases over years and this effect was technically presented as shape parameter function of Weibull reliability function of spacecraft beyond Earth-Mars Extremity [13]. Hence technical attention is required while fabricating and validating planetary probes. JAXA's Nozomi probe is found to be the most durable and long lasted probe among all failed probes that operated for years before it encounters its malfunction.

<div align="center">**Table-1. Stages of Component Failures**</div>

| S.No | FC* | Spacecraft | Stage-1 | Stage-2 | Stage-3 | Stage-4 | Stage-5 | Stage-6 | Stage-7 |
|---|---|---|---|---|---|---|---|---|---|
| 1. | LVM | 1M No.1 | Gyroscope | Attitude Control System | Horizon Sensor | | | | |
| 2. | LVM | 2MV-4 No.1 | Turbo Pump | Ignition Engine | Booster Explosion | | | | |
| 3. | LVM | 2M No.522 | Drain Plug | Fuel Leakage $NO_2$ | Ignition Engine | | | | |
| 4. | LVM | Mariner 8 | Transistor | Electrical System | Ignition Engine | | | | |
| 5. | FIS | 1M No.2 | Oxidizer Shut-off Valve | Ignition Engine | | | | | |
| 6. | FIS | 2MV-3 No.1 | Oxidizer Pressurization | Turbo Pump | Electrical System | Ignition Engine | | | |
| 7. | FIS | Mariner 3 | Heat Shield | Solar Panel Deployment | Battery | Communication System | | | |
| 8. | FIS | 2M No.521 | Rotor | Oxidizer Pump | Ignition Engine | Booster Explosion | | | |
| 9. | FIS | Mars 96 Orbiter | Ignition Engine | | | | | | |
| 10. | FCS | Mars 3 Lander | Antenna Discharge | Communication System | | | | | |
| 11. | FCS | Mars 6 Lander | Transistor | Onboard Computer | Communication System | | | | |
| 12. | FCS | Mars Observer | Pressurization Regulator | Pyro Valve | Fuel Tank Rupture | Orientation System | Solar Panels | Battery | Communication System |
| 13. | FOS | 2MV-4 No.2 | Turbo Pump | Ignition Engine | Booster Explosion | | | | |
| 14. | FOS | Phobos 2 Orbiter | Onboard Computer | Orientation System | Solar Panels | Battery | Communication System | | |
| 15. | FTCS | Zond 2 | Tug Cord | Solar Panel Deployment | Thermal Control System | Timer | Communication | | |
| 16. | FTCS | Nozomi | Electrical System | Thermal Control System | Ignition Engine | Communication System | | | |
| 17. | FOCS | Mars 4 Orbiter | Transistor | Onboard Computer | Ignition Engine | | | | |
| 18. | FOCS | Fobos-Grunt | Electronic Chips | Onboard Computer | Ignition Engine | | | | |
| 19. | FSP | MCO | Software | Attitude Control System | Communication System | | | | |
| 20. | FSP | MPL | Antenna | Communication System | Software | Retrorocket | | | |
| 21. | FSP | Schiaparelli EDM | Gyroscope | Guidance Control System | Software IMU Unit | Communication System | | | |
| 22. | FCP | Kosmos 419 Orbiter | Programming Error | Ignition Timer | Ignition Engine | | | | |
| 23. | FCP | Phobos 1 Orbiter | Programming Error | Attitude Control System | Stabilization System | Orientation System | Solar Panels | Battery | Communication System |
| 24. | FCP | Beagle 2 Lander | Programming | Communication System | | | | | |
| 25. | FDEIS | Mars 7 Lander | Transistor | Onboard Computer | Descent Engine Ignition | | | | |
| 26. | FSF | Mars 2 Lander | Spacecraft Malfunction | Parachute | | | | | |
| 27. | LBMI | Mars 2 Prop-M | Never Deployed | | | | | | |
| 28. | LBMI | Mars 3 Prop-M | Never Deployed | | | | | | |
| 29. | LBMI | Phobos 1 Lander | Never Deployed | | | | | | |
| 30. | LBMI | Phobos 2 Lander | Never Deployed | | | | | | |
| 31. | LBMI | Mars 96 Lander | Never Deployed | | | | | | |
| 32. | LBMI | Mars 96 Penetrator | Never Deployed | | | | | | |
| 33. | LBMI | Deep Space 2 | Never Deployed | | | | | | |
| 34. | LBMI | Fobos-Grunt Lander | Spacecraft Destroyed | | | | | | |
| 35. | LBMI | Yinghuo-1 | Spacecraft Destroyed | | | | | | |

*FC – Failure Classification

## VII. Recommendations

### A. Launch Vehicle Integration, Testing and Validation

Even though launch vehicle failure is not predominant nowadays. Recent event (Fobos-Grunt launch attempt) ensues the distrust of launch vehicle successfulness [14]. So, it is recommended that each and every component of the launch vehicle should undergo proper testing and validation prior to launcher integration. Further, like the Crew escape system, it is good to have a probe emergency escape system in advance to avoid ineffectiveness of mission effort and launch cost during failures [15].

### B. Probe Fabrication, Installation, and Validation

Recovery options for the Mars probes seem to be the most difficult task once they left for transit towards Mars. So, critical analysis, repeated testing and validation of the probe component is necessary. In addition to this, recheck and validation of probe performance and deployment systems are hardly recommended before throwing the probe into interplanetary space. Then, focusing on the section of thermal insulation and fuel management, it is significant to consider sufficient thickness of radiation suit of the spacecraft to avoid unusual changes in probe internal temperature. Because the probe, during their interplanetary transit to Mars for about 6-9 months is subjected to hazardous space radiations and solar eruptions that are capable of rupturing the probe's components. Hence, the spacecraft with good fabrication technology is highly recommended for a durable and reliable mission [16].

### C. Software Design and Programming

Modern space probes are found to have encountered software and programming issues [17-20] and it may occur at any stage of the mission due to negative space environmental impacts [21]. So, we recommend to develop and fabricate robust software. The problem concerning the programming, it can be averted with automated proofreading of computer commands prior to transmitting it to the operational probes than manual proofreading. Further, the computer program for the backup and recovery-related issues can be simulated in our ground-based laboratories and can be made available to the probes beforehand. So that the probe can execute effective recovery procedures on time without anticipate for the further command or instructions from the ground. Advanced availability is significant due to delayed communication time lag between the probe (at Mars) and the ground controllers (on Earth).

### D. Onboard Computer and Communication Guidance

The onboard computer and communication system are the principal component of the spacecraft subsystem. Even though probe computer and circuitry are protectively placed with a preferable radiation shield, computer malfunction may occur at any stage due to its long term performance over the years. So, we recommend that the probe should be installed with dual-computer and it should be capable of switching to the space computer in case of computer malfunction. Alike computer, the communication system is the sensing component of the probe to operate from a remote distance and hence we recommend to install either dual-mode communication system (i.e. dual antenna) or anti-environmental proof antennas that should resist against the hostile planetary environmental condition.

### E. Recovery Preferences

Recovery procedures can be only applied to the probe stranded in low-earth orbit. In the case of temporary probe failure, an effective recovery method should be made available in advance and executed. Besides, it is surprising to see that the degradation period of the probe is fairly greater than its mission life [22]. This circumstance explains that the probe subsists even after their defunct. So, it is desirable to find a possible recovery option for the probe stranded in LEO. Identically for the probe beyond low-earth orbit or the probe that approached Mars, it can be redirected to any safest location (either to the moons of Mars or into the Mars) for any feasible recovery option in future.

## VIII. Conclusions

Mars is the only accessible planet for life in our solar system after our Earth. Space Industries are striving towards the red planet to explore and answer the question to the origin and evolution of life. Since 1960, scientists and explorers have been spuriously exploring Mars with more than 70 spacecrafts (Overall Mission Summary is shown in Fig 13) and many have future insight for the human-crewed mission. But many spacecrafts have suffered technical issues and setbacks before reaching Mars or accomplishing its mission. Hence, the current state of Mars endeavors demands a failure analysis for progressing towards a sustainable mission. In the view of this ultimatum, we conducted a study and analysis on failed Mars probes. That resulted in the careful analysis and study performed on 33 different Mars probes. In our study, the failure modes of various probes were classified and characterized to show the first source of failure that triggers recurrent failures. Then the series of failures were distinguished as the stages of component failures. In addition to this, some useful parameters of probes were comprehensively shown in graphical representations and tables. Further, subject to the modes of failure, pertinent recommendations were discussed along with possible recovery preferences. Finally, we expect that our past report [1] and the current study may provide some useful framework to the Mars explorers to stride towards a progressive and effective mission.

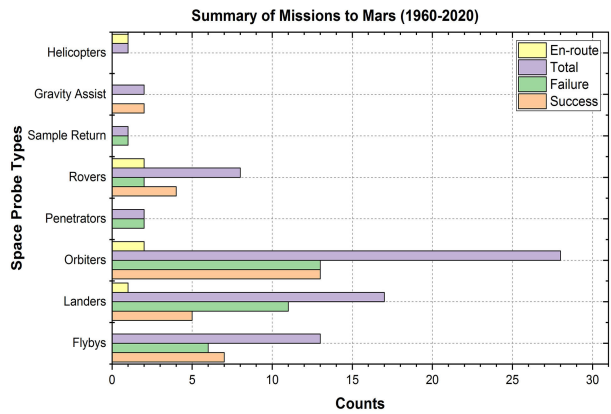

Fig 13 Summary of Missions to Mars from 1960s to 2020s

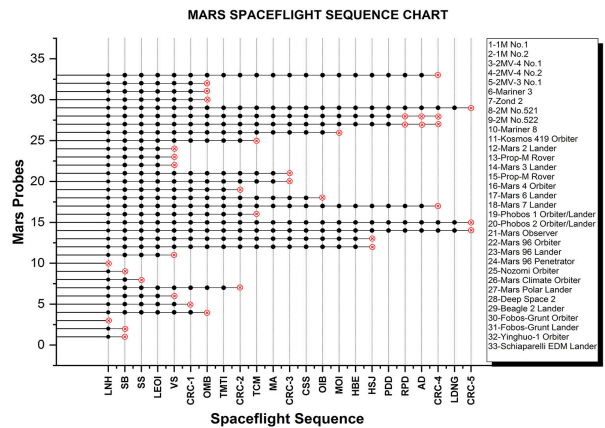

Fig 14 Mars Spaceflight Sequence Chart [1]

## Acknowledgements

I (Malaya Kumar Biswal) would like to thank my supervisor, **Prof. A. Ramesh Naidu** (author), for the patient guidance, encouragement and advice he has provided throughout my time as his student. I have been extremely lucky to have a supervisor who cared so much about my work, and who responded to my queries so promptly. Further I would like to extend my sincere thankfulness to all of my lovable friends for their financial support for the conference participation.

## Dedication

I (Malaya Kumar Biswal) would like to dedicate this work to my beloved mother late. **Mrs. Malathi Biswal** for her motivational speech and emotional support throughout my life.

## Conflict of Interest

The authors have no conflict of interest to report.

## Supplementary Resources

**Mars Missions Failure Report Assortment: Review and Conspectus**. Published by American Institute of Aeronautics and Astronautics in the Conference Proceedings of 2020 AIAA Propulsion and Energy Forum. Accessible at https://doi.org/10.2514/6.2020-3541

# Appendices

## Numerical Data for A Study on Mars Probe Failures

**Classifications of Failures:**

1) LVM    -    Launch Vehicle Malfunction
2) FIS    -    Failure of Ignition System
3) FCS    -    Failure of Communication System
4) FOS    -    Failure of Orientation System
5) FTCS  -    Failure of Thermal Control System
6) FOCS  -    Failure of Onboard Computer System
7) FSP    -    Failure of Software Program
8) FCP    -    Failure in Computer Programming
9) FDEIS    -    Failure of Descent Engine Ignition System
10) FSF   -    Failure of Spacecraft Function
11) LBMI -    Lost Before Mission Initiation

| LVM – Launch Vehicle Malfunction ||||||
|---|---|---|---|---|---|
| S.No | Spacecraft | Spacecraft Mass (kg) | Launch Mass (kg) | Mission Duration (days) | Mission Degradation (days) |
| 01. | 1M.No.1 | 480 | 650 | 0.0036 | 01 |
| 02. | 2.MV-4.No.1 | 893 | 900 | 0.0034 | 01 |
| 03. | 2M.No.522 | 3800 | 4850 | 0.00048 | 01 |
| 04. | Mariner 8 | 558.8 | 997.9 | 0.0032 | 01 |
|  | **Average** | **1432.95** | **1849.475** | **0.00267** | **01** |

| FIS – Failure of Ignition System ||||||
|---|---|---|---|---|---|
| S.No | Spacecraft | Spacecraft Mass (kg) | Launch Mass (kg) | Mission Duration (days) | Mission Degradation (days) |
| 01. | 1M.No.2 | 480 | 650 | 0.0034 | 01 |
| 02. | 2.MV-3.No.1 | 890 | 900 | 0.0030 | 01 |
| 03. | Mariner 3 | 260 | 397 | 0.36 | 01 |
| 04. | 2M.No.521 | 3800 | 4850 | 0.0050 | 01 |
| 05. | Mars 96 Orbiter | 3780 | 6700 | 0.0062 | 02 |
|  | **Average** | **1842** | **2699.4** | **0.07522** | **1.20** |

| FCS – Failure of Communication System ||||||
|---|---|---|---|---|---|
| S.No | Spacecraft | Spacecraft Mass (kg) | Launch Mass (kg) | Mission Duration (days) | Mission Degradation (days) |
| 01. | Mars 3 Lander | 358 | 1210 | 188 | 188 |
| 02. | Mars 6 Lander | 635 | 3260 | 219 | 219 |
| 03. | Mars Observer | 1018 | 2565 | 331 | 330 |
|  | **Average** | **670.33** | **2345** | **246** | **245.66** |

| FOS – Failure of Orientation System ||||||
|---|---|---|---|---|---|
| S.No | Spacecraft | Spacecraft Mass (kg) | Launch Mass (kg) | Mission Duration (days) | Mission Degradation (days) |
| 01. | 2.MV-4.No.2 | 893 | 900 | 140 | 230 |
| 02. | Phobos 2 Orbiter | 2420 | 6220 | 258 | 258 |
|  | **Average** | **1656.5** | **3560** | **199** | **244** |

| FTCS - Failure of Thermal Control System ||||||
|---|---|---|---|---|---|
| S.No | Spacecraft | Spacecraft Mass (kg) | Launch Mass (kg) | Mission Duration (days) | Mission Degradation (days) |
| 01. | Zond 2 | 890 | 996 | 18 | 249 |
| 02. | Nozomi | 258 | 356 | 1983 | 1985 |
|  | **Average** | **574** | **676** | **1000.5** | **1117** |

| FOCS - Failure of Onboard Computer System ||||||
|---|---|---|---|---|---|
| S.No | Spacecraft | Spacecraft Mass (kg) | Launch Mass (kg) | Mission Duration (days) | Mission Degradation (days) |
| 01. | Mars 4 Orbiter | 2265 | 3440 | 09 | 204 |
| 02. | Fobos-Grunt | 1560 | 2300 | 0.00155 | 68 |
|  | **Average** | **1912.5** | **2870** | **4.5** | **136** |

| FSP – Failure of Software Program | | | | |
|---|---|---|---|---|
| S.No | Spacecraft | Spacecraft Mass (kg) | Launch Mass (kg) | Mission Duration (days) | Mission Degradation (days) |
| 01. | MCO | 338 | 638 | 286 | 286 |
| 02. | MPL | 290 | 583 | 334 | 334 |
| 03. | Schiaparelli EDM | 280 | 600 | 03 | 219 |
| | **Average** | **302.66** | **607** | **207.66** | **279.66** |

| FCP – Failure in Computer Programming | | | | |
|---|---|---|---|---|
| S.No | Spacecraft | Spacecraft Mass (kg) | Launch Mass (kg) | Mission Duration (days) | Mission Degradation (days) |
| 01. | Kosmos 419 | 4549 | 4549 | 0.0625 | 02 |
| 02. | Phobos-1 Orbiter | 2420 | 6220 | 52 | 119 |
| 03. | Beagle-2 Lander | 09 | 33.2 | 183 | 206 |
| | **Average** | **2326** | **3600.73** | **78.354** | **109** |

| FDEIS - Failure of Descent Engine Ignition System | | | | |
|---|---|---|---|---|
| S.No | Spacecraft | Spacecraft Mass (kg) | Launch Mass (kg) | Mission Duration (days) | Mission Degradation (days) |
| 01. | Mars 7 Lander | 635 | 3260 | 214 | 212 |
| | **Average** | **635** | **3260** | **214** | **212** |

| FSF - Failure of Spacecraft Function | | | | |
|---|---|---|---|---|
| S.No | Spacecraft | Spacecraft Mass (kg) | Launch Mass (kg) | Mission Duration (days) | Mission Degradation (days) |
| 01. | Mars 2 Lander | 358 | 1210 | 192 | 192 |
| | **Average** | **358** | **1210** | **192** | **192** |

| LBMI – Lost Before Mission Initiation | | | | |
|---|---|---|---|---|
| S.No | Spacecraft | Spacecraft Mass (kg) | Launch Mass (kg) | Mission Duration (days) | Mission Degradation (days) |
| 01. | Mars 2 Prop-M | 4.5 | 4.5 | - | 192 |
| 02. | Mars 3 Prop-M | 4.5 | 4.5 | - | 188 |
| 03. | Phobos 1 Lander | 570 | 3800 | - | 119 |
| 04. | Phobos 2 Lander | 570 | 3800 | - | 277 |
| 05. | Mars96 Lander | 75 | 75 | - | 02 |
| 06. | Mars96 Penetrator | 120 | 120 | - | 02 |
| 07. | Deep Space 2 | 2.4 | 04 | - | 334 |
| 08. | Fobos-G Lander | 106 | 296 | - | 68 |
| 09. | Yinghuo-1 | 115 | 115 | - | 68 |
| | **Average** | **174.155** | **913.22** | **0** | **138.88** |

| Rate of Failures by Classifications | | | |
|---|---|---|---|
| S.No | Classifications | No. of Space Probes | Rate of Failure (%) |
| 01. | LVM | 04 | 11.42 |
| 02. | FIS | 05 | 14.28 |
| 03. | FCS | 03 | 8.57 |
| 04. | FOS | 02 | 5.71 |
| 05. | FTCS | 02 | 5.71 |
| 06. | FOCS | 02 | 5.71 |
| 07. | FSP | 03 | 8.57 |
| 08. | FCP | 03 | 8.57 |
| 09. | FDEIS | 01 | 2.86 |
| 10. | FSF | 01 | 2.86 |
| 11. | LBMI | 09 | 25.71 |
| | **Total** | **35** | **99.97** |

| Failure Rate by Spacecraft Type | | | |
|---|---|---|---|
| S.No | Spacecraft Type | NOS | Failure Rate (%) |
| 01. | Flybys | 06 | 17.14 |
| 02. | Landers | 11 | 31.43 |
| 03. | Orbiters | 13 | 37.14 |
| 04. | Penetrators | 02 | 5.71 |
| 05. | Rovers | 02 | 5.71 |
| 06. | Sample Return | 01 | 2.86 |
| | Total | 35 | 99.99 |

| Mars Mission Summary by Spacecraft Type | | | | |
|---|---|---|---|---|
| S.No | Spacecraft Type | Success | Failure | Total |
| 01. | Flybys | 07 | 06 | 13 |
| 02. | Landers | 05 | 11 | 16 |
| 03. | Orbiters | 13 | 13 | 26 |
| 04. | Penetrators | - | 02 | 02 |
| 05. | Rovers | 04 | 02 | 06 |
| 06. | Sample Return | - | 01 | 01 |
| 07. | Gravity Assist | 02 | - | 02 |
| | Total | 31 | 35 | 66 |

| Failure Rate by Country | | | |
|---|---|---|---|
| S.No | Country | NOS | Failure Rate (%) |
| 01. | CHINA | 01 | 2.86 |
| 02. | EUROPE | 02 | 5.71 |
| 03. | JAPAN | 01 | 2.86 |
| 04. | USA | 06 | 17.14 |
| 05. | USSR | 25 | 71.43 |
| | Total | 35 | 100 |

| Failure Rate by Launch Vehicle | | | |
|---|---|---|---|
| S.No | Launch Vehicle | NOS | Failure Rate (%) |
| 01. | Atlas | 01 | 7.14 |
| 02. | Molniya | 04 | 28.57 |
| 03. | Proton | 06 | 42.85 |
| 04. | Zenit | 03 | 21.42 |
| | Total | 14 | 99.98 |

| Comparison Average Mass, Launch Mass, Duration and Degradation by Failure Classifications | | | | | |
|---|---|---|---|---|---|
| S.No | Classifications | Average Spacecraft Mass (kg) | Average Launch Mass (kg) | Average Mission Duration (days) | Average Mission Degradation (days) |
| 01. | LVM | 1432.95 | 1849.475 | 0.00267 | 1 |
| 02. | FIS | 1842 | 2699.4 | 0.07552 | 1.20 |
| 03. | FCS | 670.33 | 2345 | 246 | 245.66 |
| 04. | FOS | 1656.5 | 3560 | 199 | 244 |
| 05. | FTCS | 574 | 676 | 1000.5 | 1117 |
| 06. | FOCS | 1912.5 | 2870 | 4.5 | 136 |
| 07. | FSP | 302.66 | 607 | 207.66 | 279.66 |
| 08. | FCP | 2326 | 3600.73 | 78.354 | 109 |
| 09. | FDEIS | 635 | 3260 | 214 | 212 |
| 10. | FSF | 358 | 1210 | 192 | 192 |
| 11. | LBMI | 174.155 | 913.22 | 0 | 138.88 |
| | Total | 11884.095 | 23590.825 | 2142.09219 | 2676.4 |

| Spacecraft's Failed Components Counts & Rate of Failure (%) ||||||
|---|---|---|---|---|---|
| S.No | Components | Failure Rate % | S.No | Components | Counts |
| 01. | Antenna | 2.33 | 01. | Antenna | 02 |
| 02. | Attitude Control System | 3.49 | 02. | Attitude Control System | 03 |
| 03. | Battery | 4.65 | 03. | Battery | 04 |
| 04. | Communication System | 13.95 | 04. | Communication System | 12 |
| 05. | Descent Engine | 1.16 | 05. | Descent Engine | 01 |
| 06. | Drain Plug | 1.16 | 06. | Drain Plug | 01 |
| 07. | Electrical System | 3.49 | 07. | Electrical System | 03 |
| 08. | Electronic Chips | 1.16 | 08. | Electronic Chips | 01 |
| 09. | Fuel Tank Rupture | 1.16 | 09. | Fuel Tank Rupture | 01 |
| 10. | Gyroscope | 2.33 | 10. | Gyroscope | 02 |
| 11. | Heat Shield | 1.16 | 11. | Heat Shield | 01 |
| 12. | Horizon Sensor | 1.16 | 12. | Horizon Sensor | 01 |
| 13. | Ignition Engine | 13.95 | 13. | Ignition Engine | 12 |
| 14. | Fuel (NO2) Leakage | 1.16 | 14. | Fuel (NO2) Leakage | 01 |
| 15. | Onboard Computer | 5.81 | 15. | Onboard Computer | 05 |
| 16. | Orientation Control System | 3.49 | 16. | Orientation Control System | 03 |
| 17. | Oxidizer Pump | 1.16 | 17. | Oxidizer Pump | 01 |
| 18. | Oxidizer Shut-off Valve | 1.16 | 18. | Oxidizer Shut-off Valve | 01 |
| 19. | Parachute | 1.16 | 19. | Parachute | 01 |
| 20. | Pressurization Regulator | 1.16 | 20. | Pressurization Regulator | 01 |
| 21. | Pressurization System | 1.16 | 21. | Pressurization System | 01 |
| 22. | Programming | 3.49 | 22. | Programming | 03 |
| 23. | Pyro Valve | 1.16 | 23. | Pyro Valve | 01 |
| 24. | Retrorocket | 1.16 | 24. | Retrorocket | 01 |
| 25. | Rotor | 1.16 | 25. | Rotor | 01 |
| 26. | Software | 3.49 | 26. | Software | 03 |
| 27. | Solar Panel | 5.81 | 27. | Solar Panel | 05 |
| 28. | Spacecraft | 1.16 | 28. | Spacecraft | 01 |
| 29. | Stabilization System | 1.16 | 29. | Stabilization System | 01 |
| 30. | Thermal Control System | 2.33 | 30. | Thermal Control System | 02 |
| 31. | Timer | 2.33 | 31. | Timer | 02 |
| 32. | Transistor | 4.65 | 32. | Transistor | 04 |
| 33. | Tug Cord | 1.16 | 33. | Tug Cord | 01 |
| 34. | Turbo Pump | 3.49 | 34. | Turbo Pump | 03 |
| | Total | 100 | | Total | 86 |

| Overall Mars Missions Summary (1960 – 2020) ||||||||
|---|---|---|---|---|---|---|---|
| S.No | Spacecraft Type | 1960s | 1970s | 1980s | 1990s | 2000s | 2010s | Total |
| 01. | Flybys | 9 | 2 | - | - | - | 2 | 13 |
| 02. | Landers | 1 | 6 | 2 | 3 | 2 | 2 | 16 |
| 03. | Orbiters | 2 | 9 | 2 | 5 | 3 | 5 | 26 |
| 04. | Penetrators | - | - | - | 2 | - | - | 2 |
| 05. | Rovers | - | 2 | - | 1 | 2 | 1 | 6 |
| 06. | Sample Return | - | - | - | - | - | 1 | 1 |
| 07. | Gravity Assist | - | - | - | - | 2 | - | 2 |
| | Total | 12 | 19 | 4 | 11 | 9 | 11 | 66 |

| Time Duration : 1960s – 1990s | | |
|---|---|---|
| First Source Encounter Duration After Launch | | |
| S.No | Duration | Rate (%) |
| 01. | 0 - 1 Months | 55.56 |
| 02. | 1 - 2 Months | 5.56 |
| 03. | 4 - 5 Months | 16.67 |
| 04. | 5 - 6 Months | 16.67 |
| 05. | 6 - 8 Months | 5.56 |
| | Total | 100 |

| Time Duration : 1990s – 2020s | | |
|---|---|---|
| First Source Encounter Duration After Launch | | |
| S.No | Duration | Rate (%) |
| 01. | 0 - 1 Months | 25 |
| 02. | 5 - 6 Months | 25 |
| 03. | 7 - 8 Months | 37.5 |
| 04. | 1 - 4 Years | 12.5 |
| | Total | 100 |

| Power Distribution of Failed Mars Probes | | |
|---|---|---|
| S.No | Power | Rate (%) |
| 01. | 0 - 100 W | 11.428 |
| 02. | 100 - 500 W | 14.285 |
| 03. | 500 - 1000W | 2.857 |
| 04. | 1000 - 1200 W | 2.857 |
| 05. | Unknown | 68.571 |
| | Total | 100 |

| Failed Mars Probe Parameters | | | | | | | |
|---|---|---|---|---|---|---|---|
| S.No | Mars Probe | Decay Date | Probe Mass | Launch Mass | Mission Duration | Mission Degradation | Power |
| Units | | | kilograms | kilograms | Days | Days | watts |
| 01. | 1M No.1 Flyby | 10-Oct-1960 | 480 | 650 | 0.0036 | 01 | - |
| 02. | 1M No.2 Flyby | 14-Oct-1960 | 480 | 650 | 0.0034 | 01 | - |
| 03. | 2 MV-4 No.1 Flyby | 26-Feb-1963 | 893 | 900 | 0.0034 | 125 | - |
| 04. | 2 MV-4 No.2 Flyby | 19-Jun-1963 | 893 | 900 | 140 | 230 | - |
| 05. | 2 MV-3 No.1 Lander | 19-Jan-1963 | 890 | 900 | 0.0030 | 227 | - |
| 06. | Mariner 3 Flyby | 06-Nov-1964 | 260 | 397 | 0.36 | 01 | 300 |
| 07. | Zond 2 Flyby | 18-Dec-1964 | 890 | 996 | 18 | 249 | - |
| 08. | 2M No.521 Orbiter | 27-Mar-1969 | 3800 | 4850 | 0.0050 | 01 | - |
| 09. | 2M No.522 Orbiter | 02-Apr-1969 | 3800 | 4850 | 0.00048 | 01 | - |
| 10. | Mariner 8 Orbiter | 09-May-1971 | 558.8 | 997.9 | 0.0032 | 01 | 500 |
| 11. | Kosmos 419 Orbiter | 12-May-1971 | 4549 | - | 0.0625 | 02 | - |
| 12. | Mars 2 (M-71) Lander | 27-Nov-1971 | 358 | 1210 | 192 | 192 | - |
| 13. | Prop-M Rover | 27-Nov-1971 | 4.5 | - | - | 192 | - |
| 14. | Mars 3 (M-71) Lander | 02-Dec-1971 | 358 | 1210 | - | 188 | - |
| 15. | Prop-M Rover | 02-Dec-1971 | 4.5 | - | - | 188 | - |
| 16. | Mars 4 Orbiter | 10-Feb-1974 | 2265 | 3440 | 09 | 204 | - |
| 17. | Mars 6 Lander | 12-Mar-1974 | 635 | 3260 | 219 | 219 | - |
| 18. | Mars 7 Lander | 09-Mar-1974 | 635 | 3260 | 214 | 212 | - |
| 19. | Phobos 1 Orbiter | 03-Nov-1988 | 2420 | 6220 | - | 119 | - |
| 20. | Phobos 1 Lander | 03-Nov-1988 | 570 | 3800 | - | 119 | - |
| 21 | Phobos 2 Orbiter | 27-Mar-1989 | 2420 | 6220 | 258 | 258 | - |
| 22. | Phobos 2 Lander | 15-Apr-1989 | 570 | 3800 | - | 277 | - |
| 23. | Mars Observer Orbiter | 21-Aug-1993 | 1018 | 2565 | 331 | 330 | 1147 |
| 24. | Mars 96 Orbiter | 18-Nov-1996 | 3780 | 6700 | 0.0062 | 02 | - |
| 25. | Mars 96 Lander | 18-Nov-1996 | 75 | - | - | 02 | 35 |
| 26. | Mars 96 Penetrator | 18-Nov-1996 | 120 | - | - | 02 | 15 |
| 27. | Nozomi Orbiter | 09-Dec-2003 | 258 | 356 | 1983 | 1985 | - |
| 28. | MCO Orbiter | 23-Sep-1999 | 338 | 638 | 286 | 286 | 500 |
| 29. | MPL Lander | 03-Dec-1999 | 290 | 583 | 334 | 334 | 200 |
| 30. | Deep Space 2 Penetrator | 03-Dec-1999 | 2.4 | - | 334 | 334 | - |
| 31. | Beagle 2 Lander | 25-Dec-2003 | 09 | 33.2 | 183 | 206 | 60 |
| 32. | Fobos-Grunt Orbiter | 15-Jan-2012 | 1560 | 2300 | 0.00155 | 68 | 1000 |
| 33. | Fobos-Grunt Sample | 15-Jan-2012 | 106 | 296 | - | 68 | 300 |
| 34. | Yinguo-1 Orbiter | 15-Jan-2012 | 115 | - | - | 68 | 90 |
| 35. | Schiaparelli EDM Lander | 19-Oct-2016 | 280 | 600 | 03 | 219 | - |

American Institute of Aeronautics and Astronautics